\documentclass[referee,a4paper,11pt,traditabstract]{swsc} 

\usepackage{amssymb,amsbsy,amsmath}
\usepackage{graphicx}
\usepackage{txfonts}
\usepackage[authoryear,round]{natbib}
\usepackage[backref]{hyperref}
\usepackage{url}

\hypersetup{colorlinks=true,citecolor=cyan,urlcolor=cyan,linkcolor=blue}

\renewcommand{\aap}{Astronomy and Astrophysics}
\renewcommand{\apjl}{Astrophysical Journal Letters}

\begin{document}

\title{A Center-Median Filtering Method for Detection of Temporal Variation in Coronal Images}
\titlerunning{Coronal RCM Filtering \& Detector Statistics}
\author{Joseph Plowman\inst{1}}
\authorrunning{Joseph Plowman}
\institute{High Altitude Observatory, National Center for Atmospheric Research\footnote{The National Center for Atmospheric Research is sponsored by the National Science Foundation.}, P.O. Box 3000, Boulder, CO 80307-3000, USA\\
\email{\href{mailto:plowman@ucar.edu}{plowman@ucar.edu}}}
\abstract{Events in the solar corona are often widely separated in their timescales, which can allow them to be identified when they would otherwise be confused with emission from other sources in the corona. Methods for cleanly separating such events based on their timescales are thus desirable for research in the field. This paper develops a technique for identifying time-varying signals in solar coronal image sequences which is based on a per-pixel running median filter and an understanding of photon-counting statistics. Example applications to `EIT Waves' (named after EIT, the EUV Imaging Telescope on the Solar and Heliospheric Observatory) and small-scale dynamics are shown, both using 193 \AA\ data from the Atmospheric Imaging Assembly (AIA) on the Solar Dynamics Observatory. The technique is found to discriminate EIT Waves more cleanly than the running and base difference techniques most commonly used. It is also demonstrated that there is more signal in the data than is commonly appreciated, finding that the waves can be traced to the edge of the AIA field of view when the data are rebinned to increase the signal-to-noise ratio.}
\maketitle
\section{Introduction}\label{sec:intro}

The solar corona possesses a wide variety of observed timescales, ranging from rapid brightenings which occur over a few seconds, to solar flares with timescales of minutes, to active region evolution over days, and so on. Often, the timescales of events of interest are widely separated, which can allow them to be identified when they would otherwise be lost in the background emission from other sources in the corona. Interesting phenomena include large scale `EIT' waves \citep[e.g.,][discussed later in this section]{Dereetal97} and small scale microflares and nanoflares whose energy distribution pertains to the heating of the corona \citep[e.g.,][]{parnelljupp_apj2000}. Methods of temporal filtering to separate these timescales are thus of considerable interest in solar physics.

One commonly-used method for separating these timescales are `running difference' images \citep[e.g.,][]{Thompsonetal99}, which are simply the difference between two adjacent frames in an image sequence. This has the advantage of being straightforward to implement and it has a clear mathematical definition (it is a time derivative, modulo the sampling cadence). However, running differences are sensitive to a wide variety of signals, which can make them difficult to interpret. For example, they will show a temporary brightening as a positive signal (during its rising phase) followed by a negative signal (as the brightness decays to the background level). As a result, it is not clear where the brightening ends, or if it is accompanied by a dimming relative to the pre-brightening emission level\footnote{The term `dimming' in this paper is used to refer to episodes of reduced intensity in emission images in general, not necessarily to a dimming caused by (for instance) a depletion of coronal material.} \citep[this limitation is alluded to in][]{AttrillWillsDavey10, Thompsonetal99}.

Another commonly used method which partially avoids these issues is that of base difference (and percentage base difference) images \citep[e.g.,][]{Longetal14}. This simply picks a frame in the image sequence (usually an early one) to use as a reference image and computes the differences between it and each other frame. Unlike the running difference, it will show whether or not the image returns to the background level (i.e., that of the reference image). However, it is sensitive to any drift of the image with respect to the reference: Over time the base difference images will become dominated by long-term drift in the sequence, reducing the ability to identify smaller, more rapid variations \citep[see discussion in Section 3 of][]{WillsDavey06}. Other variations sometimes found are the base or running ratios (related to percentage difference images via a unit offset), as found in \cite{Liu_etal_apj2012,Downs_etal_apj2011}, but this paper restricts its comparisons to base and running differences.

The method described here is a `running center-median' (RCM) filter for resolving dynamics in the corona. Like the running difference and base difference methods, it computes a difference for each image in the sequence. In this case, the difference is between the current image and the per-pixel median of all images falling within a specified time interval of the current image (e.g., within $\pm 10$ minutes). Using the median in this way avoids the limitations of the running difference and base difference methods described above: it shows a temporary brightening as such (rather than as a brightening followed by a dimming), and it adapts to long-term drift in the image sequence. In addition, the use of the median in particular has the property of being insensitive to monotonic trends in the window. 

Although median filtering is widely known in solar physics \citep[and other disciplines; it is covered in textbooks such as][]{GonzalezWoods}, it is not well represented in the literature, and does not appear to be used in the way described here. Instead, most \citep[e.g.][]{Aschwandenetal2012} use the median to summarize the central tendencies of some population they have measured (e.g., current densities in coronal loops, speeds of coronal mass ejections, etc). A relatively small fraction deal with median filtering, but most of those use it for smoothing or noise reduction: \cite{Fuhrmannetal2007} mention it as a way of smoothing noisy magnetograph data; \cite{Jingetal2000} use a 5 month temporal median filter to measure long-term background coronal emission; \cite{Glukhov1997} uses a spatial median to remove `salt and pepper' noise from SXT data; and \cite{Iwaietal2013} apply a median in frequency to remove interference from radio telescope data.  

Of the papers found in the literature, those which use the median for enhancement of smaller-scale or dynamic features include \cite{Duncan1983}, which considers spatial median filtering for enhancement of coronagraph images, and \cite{Terzoetal2011}, which compares the medians of a (per-pixel) time series to the corresponding averages for Hinode X-Ray Telescope (XRT) data. This last is the most similar method found in the literature, but differs in some notable ways from the method presented here. These differences will be considered in Section \ref{sec:discussion}, after the RCM method has been explained and demonstrated.

The method is demonstrated in Section \ref{sec:examples}, with the primary example being a coronal `EIT wave'. These are faint, global-scale disturbances observed in conjunction with solar flares and coronal mass ejections (CMEs). They were first observed in the late 1990s \citep[e.g.,][]{Dereetal97, Thompsonetal99} by the Extreme Ultraviolet Imaging Telescope (EIT) on the Solar and Heliospheric Observatory (SOHO) spacecraft. The nature of these waves is a topic of active research, with some researchers favoring a magnetosonic origin \citep[as in][]{OfmanThompson_ApJ2002}, some \citep[][for instance]{Aschwanden_AnGeo2009} favoring a large-scale reconfiguration of the coronal plasma (e.g., that it is caused by or identical with the CME), and recent developments favoring a combination of the two \citep[see][for a review]{LiuOfman14}. The waves are difficult to perceive in intensity images, and are typically detected using base or running difference images. They are therefore a good test case for a new method (such as this one) of detecting dynamic events in the corona. For additional observations and discussion see, for instance, \cite{Bieseckeretal02, WillsDaveyAttrill09, Warmuth10, LiuOfman14}. 

\section{Methodology}\label{sec:method}

The RCM filter is built around a running median, and operates pixel-by-pixel on a sequence of images. On a per-pixel basis, the core of the technique can be described as follows:

\begin{itemize}
	\item Given an intensity time series for a pixel, window the data with a specified width, $\Delta t$ (i.e., select the elements of the series falling between $t$ and $t+\Delta t$), starting at the beginning of the series.
	\item Compute the median of the intensities in the window (this is the window's `running median').
	\item Find the deviation of the window's central intensity (i.e., the one at $t+\Delta t/2$) from the median. 
	\item Assume the noise is Poisson distributed and determined by the median to compute the statistical significance of this deviation (explained below).
\end{itemize}

This is done with the same window for each pixel in the images, and an image is formed showing the statistical significance level at each pixel for the current window. The window width may be chosen to be any odd number of frames falling within the range of the time series (i.e., 11 frames would work for a 100 frame time series, but not for a 10 frame time series). An animation of the whole sequence is then made by moving the window forward one frame at a time until the end of the image sequence is reached.

\begin{figure}[!ht]
	\includegraphics[width=0.5\textwidth]{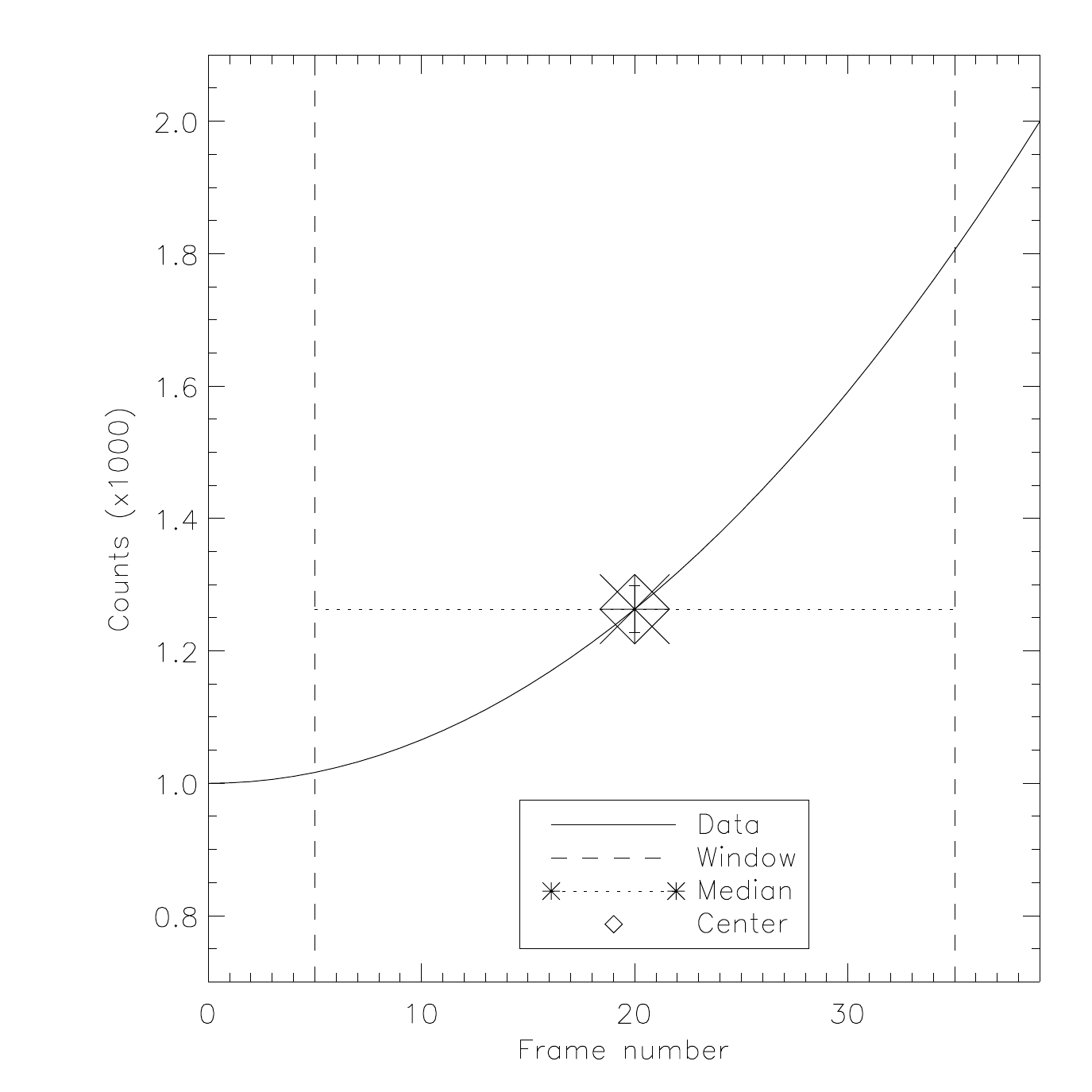}
	\includegraphics[width=0.5\textwidth]{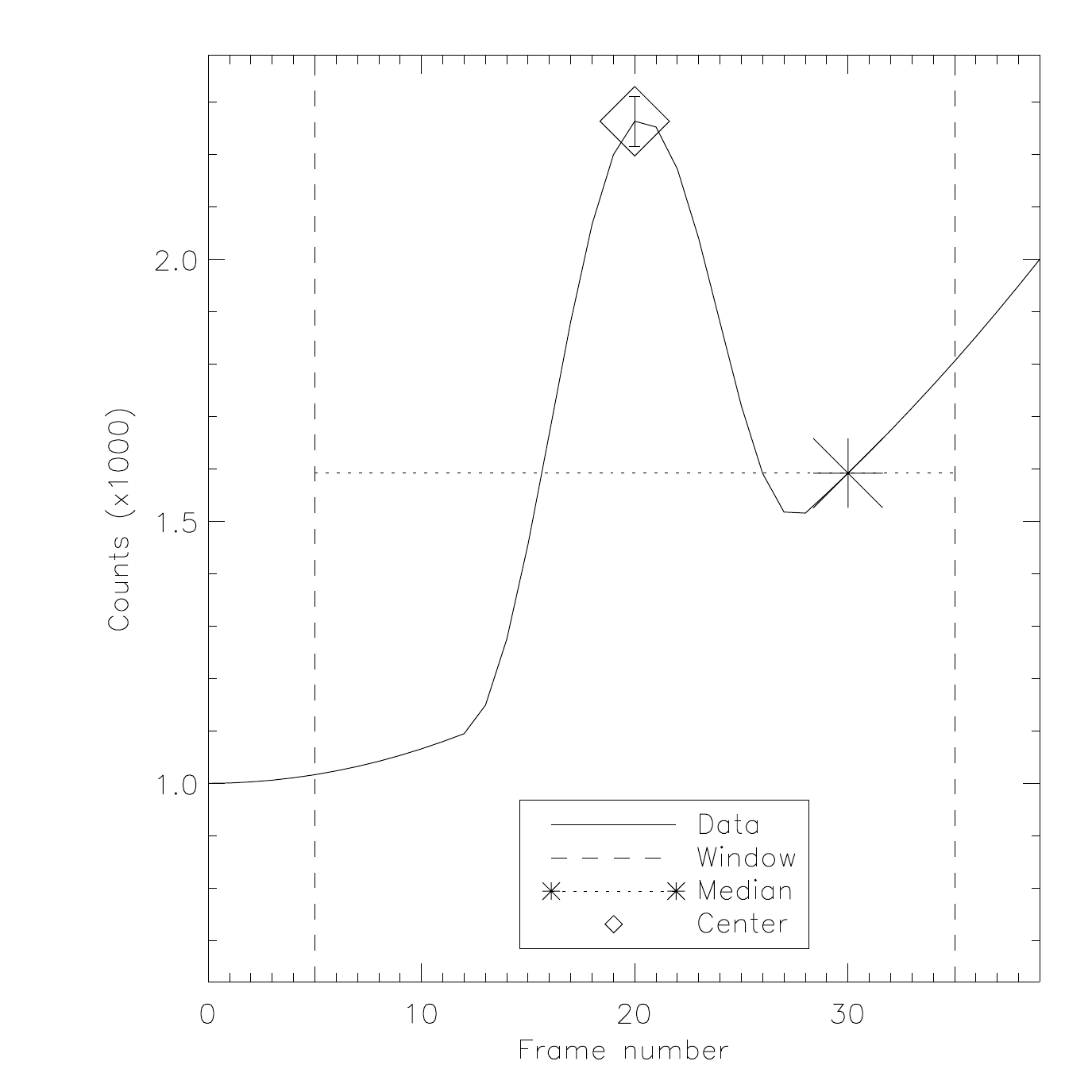}
	\caption{Left: this example illustrates the lack of response of the RCM filter to a noise-free monotonic signal; despite the significant and non-linear trend in the data, the median is equal to the center value, and the filter will return zero. Right: The response of the RCM filter to a signal with a monotonic trend and a peak. In contrast to the monotonic signal alone, there is a clear difference between the median and the center value.}\label{fig:monotonic}
\end{figure}

The filter is based on the running median because, in the low-noise limit, the central intensity will differ from the median only in the case that there is a maximum or minimum in the window (see Figure \ref{fig:monotonic}). This assumes that the window contains an odd number of frames and the sampling rate is uniform. In that case, when the window contains a monotonic time series, half of the values will be less than central value and half will be greater than it: it is therefore equal to the median, by definition. This property is not guaranteed in the case that the noise is large, however, as discussed in Section \ref{sec:limitations}.  It will also not be exactly satisfied if the window width is an even number of frames, so an odd number is recommended. 

It should be noted that effects of the solar (differential) rotation must be removed; otherwise, narrow features oriented north-south (such as coronal loops) may be erroneously detected in the resulting images. In this case, we compute coordinate corrections using the SolarSoft IDL (SSWIDL) function \texttt{diff\_rot} (off-limb coordinates are assumed to move according to their latitude) and apply them using the built-in IDL function \texttt{interpolate} with the \texttt{cubic} keyword set to -0.5\footnote{This can result in some resampling artifacts \citep[see][]{DeForest04}, but it suffices for the purpose of this proof-of-concept.}. 

The statistical significance of the window's central intensity are estimated from the Poisson cumulative distribution function, $F(C,\lambda)$, which gives the probability of obtaining some value $k$ less than or equal to the intensity (or, more properly, the photon count, $C$) as
\begin{equation}\label{poissoncdf}
	F(C,\lambda) \equiv P(k \le C|\lambda)=\sum_{k=0}^C \lambda^k e^{-\lambda}/k!,
\end{equation}
where $\lambda$ is the Poisson distribution parameter, which is set to the median in the window as described above. The `complementary' cumulative distribution function, which gives the probability of obtaining a value greater than $C$, is just
\begin{equation}\label{poissonccdf}
	\bar{F}(C,\lambda) \equiv 1-F(C,\lambda).
\end{equation}
It is straightforward to show that Equation \ref{poissonccdf} is the regularized, lower incomplete $\Gamma$ function, which can be called in IDL as \texttt{igamma(C+1,$\lambda$)}. 

For the figures in this paper (see Section \ref{sec:examples}), values of $-\log{\big[\bar{F}(C,\lambda)\big]}$ which are greater than 0.5 are shown in green; for values of $\bar{F}(C,\lambda)$ less than 0.5, $-\log{\big[F(C,\lambda)\big]}$ (i.e., the Poisson cumulative distribution function) is instead plotted in purple. Thus, green pixels indicate brightenings while purple pixels indicate dimmings, and greater image intensity indicates a more significant change in both cases.

To increase signal-to-noise ratio and reduce sensitivity to variations faster than the time scale of interest, the data may be temporally binned to a desired time interval. When combined with the RCM filter, the data is effectively band-pass filtered to time-scales longer than the binning interval but shorter than the running window width. 

The window width and binning interval are chosen by considering the time scales of the features of interest: The method is sensitive to timescales of around half the window width down to twice the frame-to-frame sampling interval (essentially the Nyquist frequency corresponding to the temporal sampling rate). In other words, the window width should be at least $\sim 2$ times the longest timescale of interest, and the temporal binning width should be half of the shortest timescale of interest. For a moving feature of thickness (along the direction of motion) $\Delta x$ and speed $v$, the duration of the window should then be at least $\sim 2\Delta x/v$. In the case of the EIT wave example in Section \ref{sec:examples}, this is approximately 50 frames at full AIA cadence, or $\sim 8$ minutes. Shorter windows will have significantly reduced sensitivity to this wave, but they will show faster variation more clearly.

\section{Examples}\label{sec:examples}

\begin{figure}[!ht]
	\includegraphics[width=0.5\textwidth]{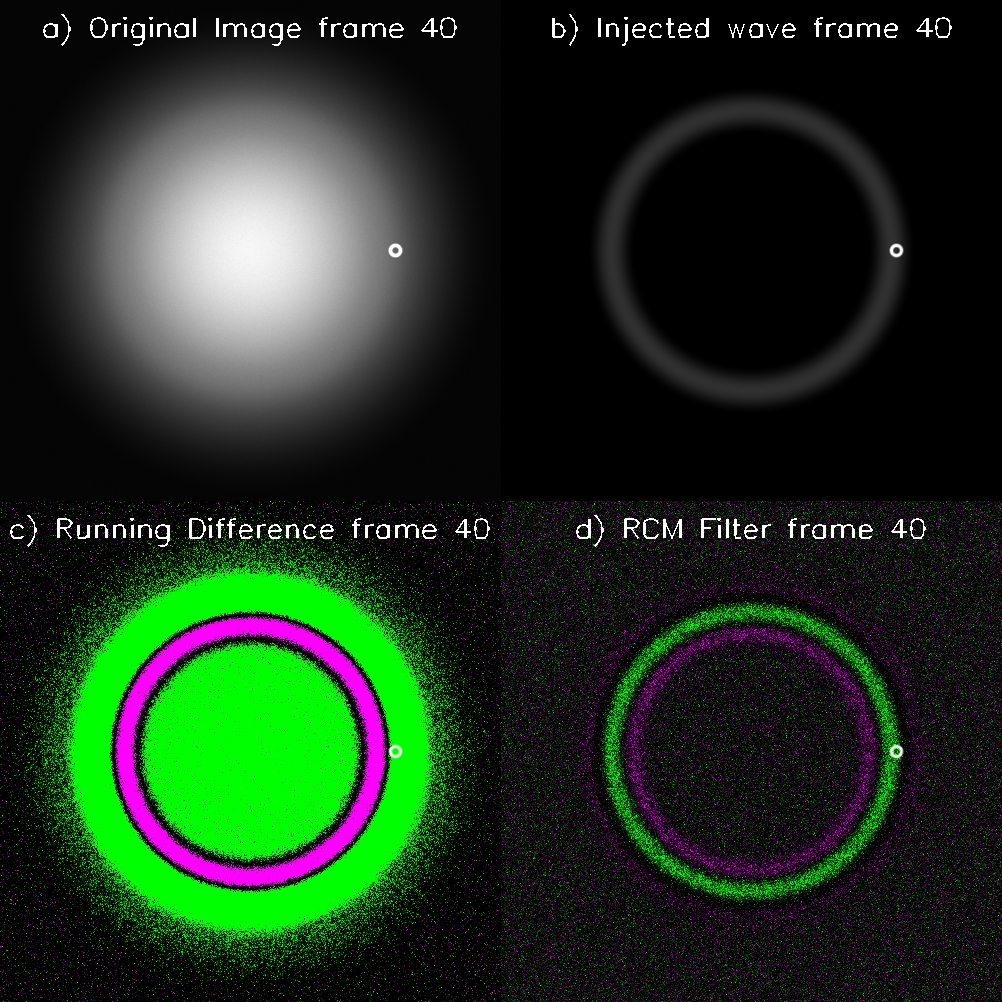}
	\includegraphics[width=0.5\textwidth]{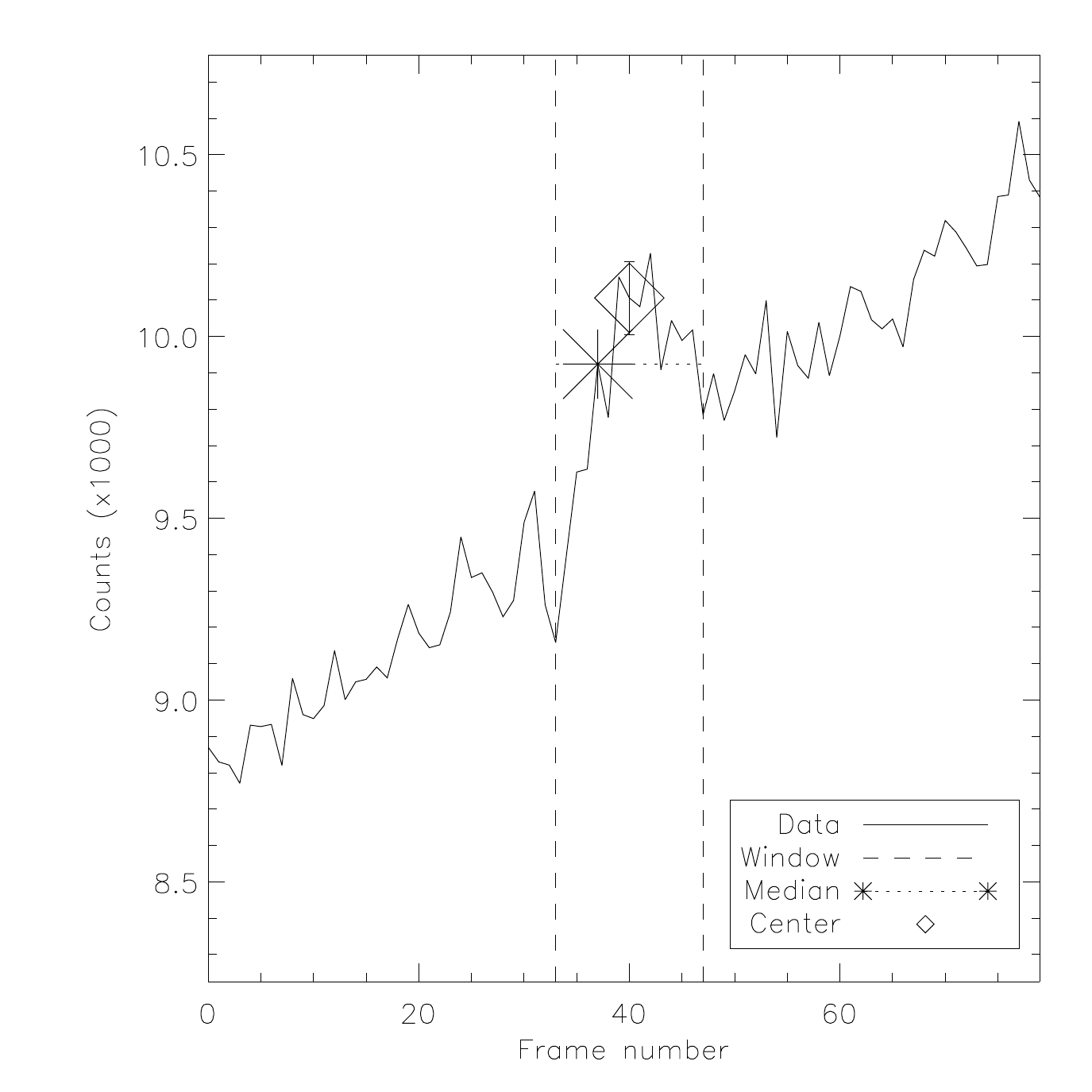}
	\caption{Left frames: Application of RCM technique to a simple test image sequence, consisting of an expanding ring-shaped `wave' superimposed on a diffuse background. (a) Shows the original frames, (b) shows the expanding ring without the background, (c) shows the running difference, and (d) shows the RCM filter. Green indicates statistically significant brightenings, purple dimmings. Right: the time series for the point indicated by the circle in the frames on the left. Readers are encouraged to download the movie showing the technique applied to the entire sequence, either from \url{https://drive.google.com/uc?id=0B8NadRgUVftLYU8tS0FCUjd1QTA&export=download} or by emailing the author.}\label{fig:poissonmap_image_test_example}
\end{figure}

Figure \ref{fig:poissonmap_image_test_example} illustrates the results of the technique applied to a test image sequence, comparing it with a running difference filter applied to the same sequence. The sequence consists of an expanding ring-shaped `wave' superimposed on a large-scale diffuse background (a circularly symmetric function which gradually brightens over the course of the sequence), and the images include the effects of shot noise. In this case, the window width was 15 frames, while the running difference subtracted the sum of the first 7 frames from that of the last 7 frames (ignoring the central frame). The right panel of Figure \ref{fig:poissonmap_image_test_example} shows the time series for a single point in the image sequence (indicated by a small circle in the images). Readers are encouraged to download the movie showing the technique applied to the entire sequence, either from \url{https://drive.google.com/uc?id=0B8NadRgUVftLYU8tS0FCUjd1QTA&export=download} or by emailing the author. Compared with the running difference, the RCM filter is not sensitive to the monotonically increasing background trend (as expected), and therefore shows the expanding ring feature much more clearly. The dimming shown by the running difference after the passage of the ring is also significantly suppressed, although not completely eliminated (the reason for this is discussed in Section \ref{sec:limitations}).

To demonstrate this technique on solar data, one example shown is an `EIT Wave' (see discussion in Section \ref{sec:intro}). This specific example occurred on April 11, 2013 around 0700 UT, and we use data from the 193 \AA\ channel of the Atmospheric Imaging Assembly (AIA) instrument on the Solar Dynamics Observatory (SDO) spacecraft \citep{Lemen_etal_SoPh2012,Pesnell_etal_SoPhSDO2012}.

\begin{figure}[!ht]
	\begin{center}
		\includegraphics[width=\textwidth]{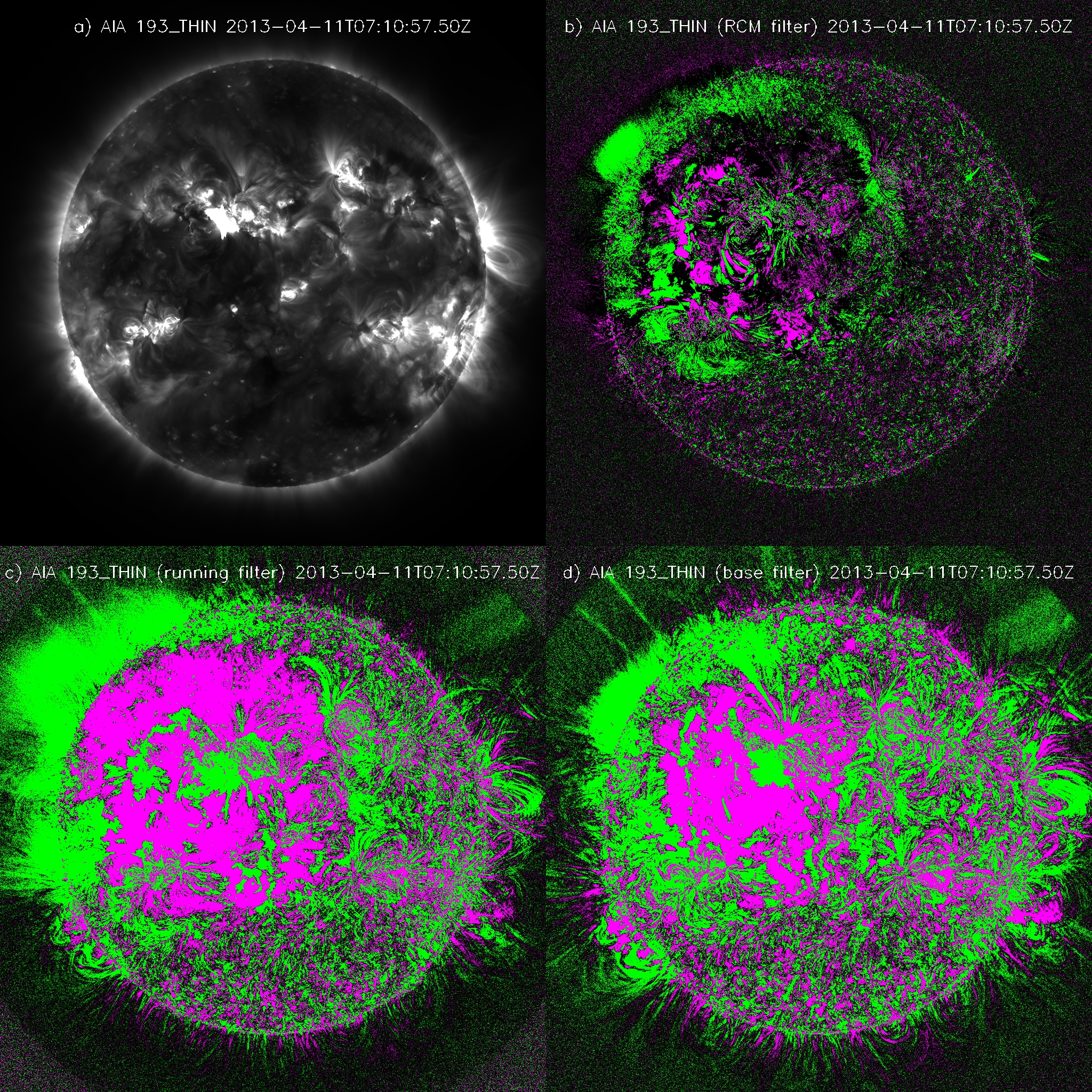}
		\includegraphics[width=\textwidth]{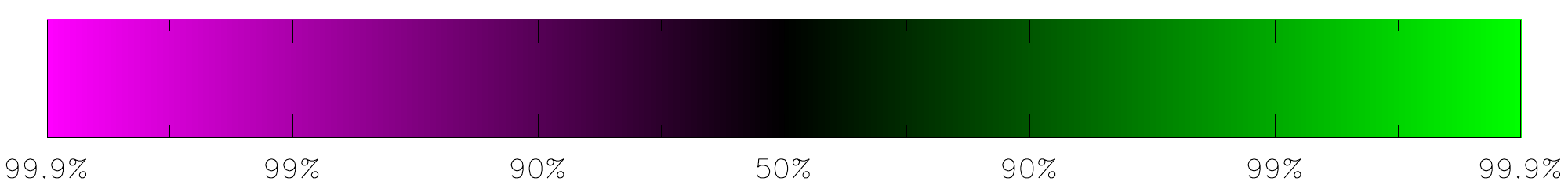}
	\end{center}
	\caption{Static frames showing (a) intensity, (b) RCM, (c) running difference, and (d) base difference filtering of an EIT wave in a AIA 193 \AA\ image sequence. The RCM filter (b) uses a 56 frame window, and each image is 4x4 pixel spatially binned. (a), (b), and (d) are 8 frame temporally binned, while (c) temporally bins the first 28 frames of the 56 frame window and subtracts them from the last 28 (binned) frames of the window. The base difference (d) uses as a base image 8 frames taken at 06:51:43.84Z. Green indicates statistically significant brightenings, purple dimmings.  Color bar shows approximate confidence corresponding to the colors in the filtered images. Readers are encouraged to download the movie showing the technique applied to the entire sequence, either from \url{https://drive.google.com/uc?id=0B8NadRgUVftLNXViZlFXOWFqLUk&export=download} or by emailing the author.}\label{fig:EITwave_long}
\end{figure}

\begin{figure}[!ht]
	\begin{center}
		\includegraphics[width=\textwidth]{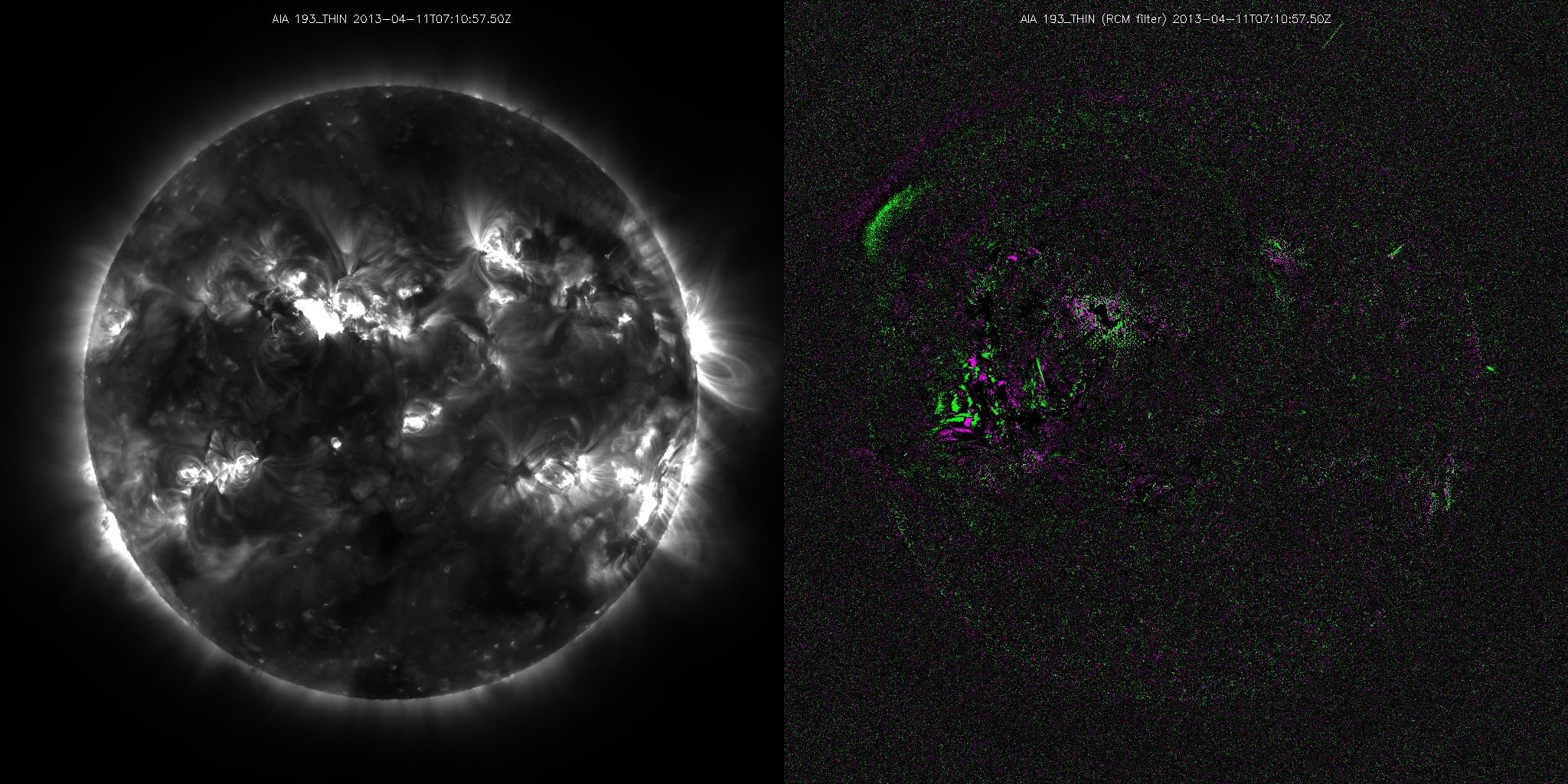}
		\includegraphics[width=\textwidth]{colorbar.pdf}
	\end{center}
	\caption{Static frame showing (right panel) RCM filtering of an AIA 193 \AA\ image sequence of an EIT wave. This plot is 4x4 pixel spatially binned, 2 frame temporally binned, and has a window width of 14 frames. Green indicates statistically significant brightenings, purple dimmings. Color bar shows approximate confidence corresponding to the colors in the image. Left panel shows the corresponding (2 frame binned) central frame.  Readers are encouraged to download the movie showing the technique applied to the entire sequence, either from \url{https://drive.google.com/uc?id=0B8NadRgUVftLTmk4R1pBbzR2dzA&export=download} or by emailing the author.}\label{fig:RCM_EITwave_short}
\end{figure}

The results of the filtering are shown in Figure \ref{fig:EITwave_long}. For the RCM filter shown in panel (b), the data has been spatially rebinned by a factor of 4 and temporally rebinned by a factor of 8 (although in the animated figures the window is moved forward by only a single AIA sampling interval, so that there are 12 seconds between each successive frame), to bring out the dim features as much as possible. The spatial binning scale is chosen to maximize sensitivity while retaining enough spatial resolution to show details of the wave (a binning factor of 8 was also tried, resulting in a 512x512 overall resolution, but the results were excessively pixelated). The temporal scale is likewise chosen to maximize sensitivity without suppressing more rapid variation. Motivated by the discussion at the end of Section \ref{sec:method}, the window width is set to 56 AIA frames (resulting in an odd number of binned frames), for a time scale range of approximately 1.5 minutes (8 AIA frames) to 10 minutes (56 AIA frames). Also considered (Figure \ref{fig:RCM_EITwave_short}) is a temporal rebinning of 2 frames (covering 12 seconds), with a window width of 14 frames (covering 168 seconds), to show rapid variation which the longer duration may have missed.

For comparison, running differences of the same data are shown in panel (c) of Figure \ref{fig:EITwave_long}. This uses the same windowing and spatial binning as the longer-duration RCM filter, and computes the difference between the sum of the frames in the first half of the window and that in the second half of the window. The motivation for this choice is to make a robust comparison with the RCM filter by using the same time window and obtaining as much SNR from the running difference as possible. In effect, it creates a pair of longer exposure images to difference, enhancing of signals in the data. In particular, signals with duration comparable to (or longer than) the window width will be most accentuated. Those which are much shorter, on the other hand, will be smoothed over, as rapid oscillations will be attenuated by the binning. For purposes of comparison with the RCM filter, the running difference images also show significance levels (rather than showing raw differences); These are computed from the differences by assuming a Gaussian distribution, with uncertainties given by the photon counts in the images being differenced. With both methods, the EIT wave is seen for much longer and further from the flare than it is visible to the eye (in the RCM sequences especially, the wave is discernable to the edge of the AIA field of view). 

The running difference images show a strong negative signal after the passage of the disturbance, but it is not clear from the running differences whether it is simply the return to some background level or evidence of a genuine propagating dimming \citep[this is mentioned in][for instance]{AttrillWillsDavey10}. The RCM filter is not as susceptible to this effect, which is related to the insensitivity to monotonic trends: consider what happens to the central value (relative to the median) as the window passes over a function with a peak and a uniform background: provided the peak occupies less than half of the window and ignoring the effects of measurement noise (this will have a similar effect to the monotonic case discussed in Section \ref{sec:limitations}), the median will be equal to the background value. Negative (purple) features in the RCM filtered images can therefore be interpreted as genuine dimmings with greater confidence.

Also shown, in panel (d) of Figure \ref{fig:EITwave_long}, is a base difference image of the same event. This uses the same spatial and temporal binning as the RCM filtered image, as well as the same color scale. The base image in this case is from 06:51:43.84Z\footnote{The filtered image sequence becomes black at this point in the time series because the current image and base image are identical (their difference is zero).} (a few minutes before the flare and EIT wave become visible), and has had the same binning applied. The wave can be discerned in the images, but the sequence becomes dominated by longer-term image drift, making it more difficult distinguish features of the wave from other accumulated variation.

Figure \ref{fig:RCM_EITwave_short} shows the same event with a temporal rebinning of 2 frames (covering 12 seconds) and a window width of 14 frames (covering 168 seconds). This high time cadence version of the filtering (Figure \ref{fig:RCM_EITwave_short}) shows features which have been missed or blurred out by the longer time interval. In particular, the wave is shown to have complex structure with multiple wavefronts not shown in the long-interval version. There are also repeated brightenings, clearly identifiable from the diffraction patterns surrounding them. These are especially noticeable in the flaring active region's core, but are present in other active regions as well (where they might be considered miniflares or microflares).

Overall, the RCM filtered images show more clearly defined features than the base or running differences, but are also noisier. In the case of the base difference, this is primarily due to accumulated variations from its longer time range. In the running difference case, there are two primary reasons: First, the running difference will tend to be larger, since it is also sensitive to monotonic trends; second, the SNR with this technique as currently implemented is limited by a single point in the window, whereas in the running difference the entire window contributes to the SNR. See Section \ref{sec:limitations} for more discussion of these noise levels.

\begin{figure}[!ht]
	\begin{center}
		\includegraphics[width=\textwidth]{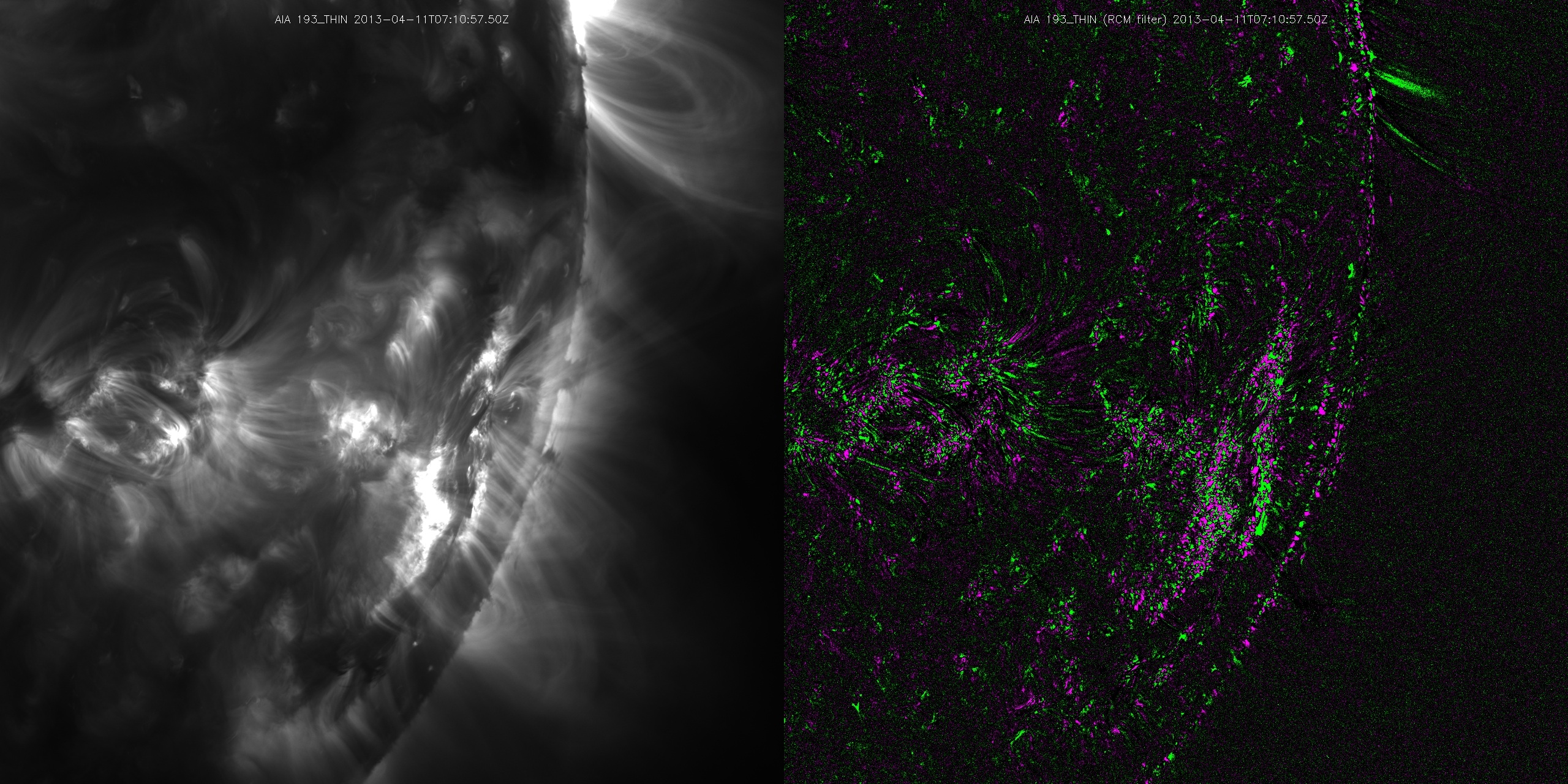}
		\includegraphics[width=\textwidth]{colorbar.pdf}
	\end{center}
	\caption{Static frame showing (right panel) RCM filtering of an AIA 193 \AA\ image sequence of the (non-flaring) lower-right quadrant, over the same time interval as the other plots. This plot is 4x4 pixel spatially binned, 8 frame temporally binned, and has a window width of 56 frames. Green indicates statistically significant brightenings, purple dimmings. Color bar shows approximate confidence corresponding to the colors in the image. Left panel shows the corresponding (8 frame binned) central frame. Readers are encouraged to download the movie showing the technique applied to the entire sequence, either from \url{https://drive.google.com/uc?id=0B8NadRgUVftLRVZ3dlVvSW1RQk0&export=download} or by emailing the author.}\label{fig:RCM_AR_long}
\end{figure}

As a second example, we show the technique applied to the same data but cropped to the non-flaring region of the sun on the lower right quadrant (Figure \ref{fig:RCM_AR_long}). This example is at full AIA temporal and spatial resolution with 8 frame binning and a window width of 56 frames. Visible are pervasive brightenings, dimmings, and motions, especially along coronal loop-like structures in active regions but also visible are smaller features in regions which might otherwise be called `quiet'. The EIT wave is also faintly visible toward the end of the sequence.

\section{Limitations}\label{sec:limitations}

\begin{figure}[!ht]
	\begin{center}
		\includegraphics[height=0.5\textheight]{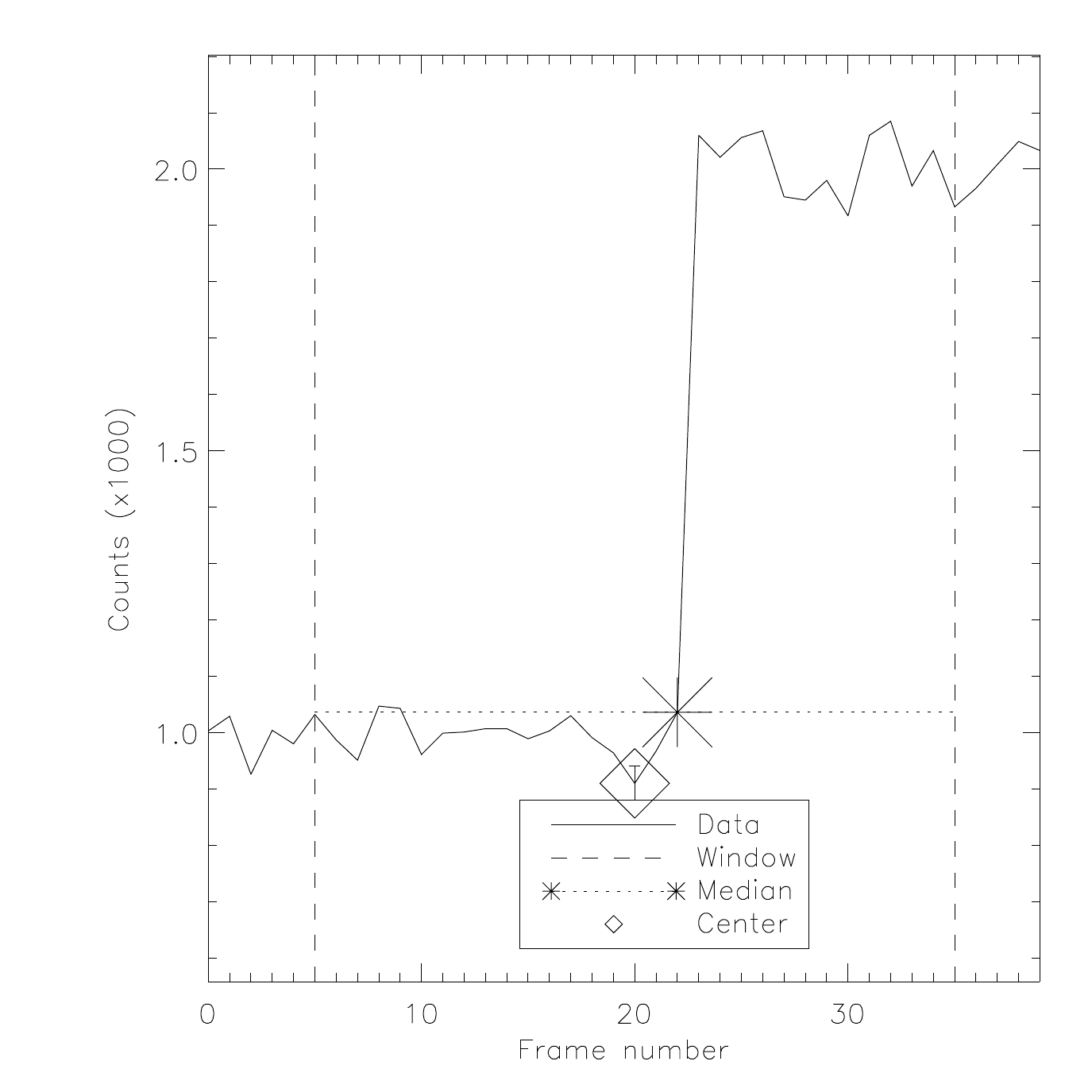}
	\end{center}
	\caption{Example illustrating response of RCM filter to a step function in presence of noise. The step causes the median to be drawn from the tail of the noise distribution of the lower part of the step. As a result, the central value will tend to be less than the median, resulting in a spurious dimming signal.}\label{fig:stepexample}
\end{figure}

The method's insensitivity to monotonic trends is not guaranteed in the presence of noise. As a concrete example consider the case where the data consists of a step function (with added noise) which begins shortly after the midpoint of the window (As shown in Figure \ref{fig:stepexample}). In this case, the upper part of the step comprises nearly half of the values in the window, and they are all much greater than those in the lower part: The so the median will then be one of the largest values in the lower part of the step, and it will therefore tend to be larger than the value at the midpoint of the window (which is a typical sample from the lower part of the step). The central value in this case will tend to indicate a dimming when none is present, and similar effects can occur for other windows which have markedly bimodal time series. The bias is because the median in the presence of noise in such cases {\em does not converge} to the same value as the median with no noise, regardless of the number of samples taken. 

The amount of bias for the step function example depends on the proximity of the step to the center point of the window. In the most extreme case, the step falls immediately before or after the center point, and the median will be either the largest or smallest value of the longer section of the step (depending on whether the shorter section is the upper or lower step, respectively). Typical deviations of this value from the noise-free median will depend on the width of the window; if the window is 19 frames wide, for example, the median will tend to fall at either the 90th or the 10th percentile ($1.3$ standard deviations), of the longer section of the step (i.e., it will either be the largest or smallest of the 10 samples drawn for that part of the step).

Another limitation is that the method is relatively insensitive compared to (for instance) the running difference, as seen in Figure \ref{fig:EITwave_long}. This is due in part to the fact that the uncertainty is limited to that in the window's midpoint, whereas in the running difference example we bin over half the window to reduce the noise. The upside to this is that the method is sensitive to transient variation from time scales of around half the window width down to the frame-to-frame scale, while the running difference works best for seeing variation at the half window width scale. This can be mitigated by reducing the width of the running difference window to a single pair of frames, but it will result in reduced sensitivity to longer-period variation and noise levels similar to the RCM filter.

Another reason for the RCM filter's relatively low sensitivity is that the median will tend to have a relatively small difference compared with any given value in the window. This is by design, however; cases where the median differs from the central frame are of interest because they correspond with some non-monotonic, (e.g., impulsive) event. This limitation can be seen in Figure \ref{fig:monotonic}; the median is higher than the monotonic background signal, so that the difference between the median and the window's middle value is smaller than the actual height of the peak. An algorithm that searches specifically for such a signal would do better in this case, but fail for signals that don't match the description (e.g., a dimming rather than a peak). The RCM method is straightforward and fairly generic, which limits its sensitivity compared to a specialized method for specific use cases.

It should also be noted that there is an ambiguity between a dimming and a peak occurring on a monotonically increasing background (or its sign-reversed counterpart, a valley on a monotonically decreasing background), which this method does not attempt to resolve. Such features will show up as a dimming (brightening in the sign-reversed case) but only when the peak falls before (after if the background is increasing) the middle of the window. This effect is illustrated in Figure \ref{fig:peakplusmonotonic}, and can be seen in the early frames of the supplemental movie corresponding to Figure \ref{fig:poissonmap_image_test_example}.

\begin{figure}[!ht]
	\begin{center}
		\includegraphics[height=0.5\textheight]{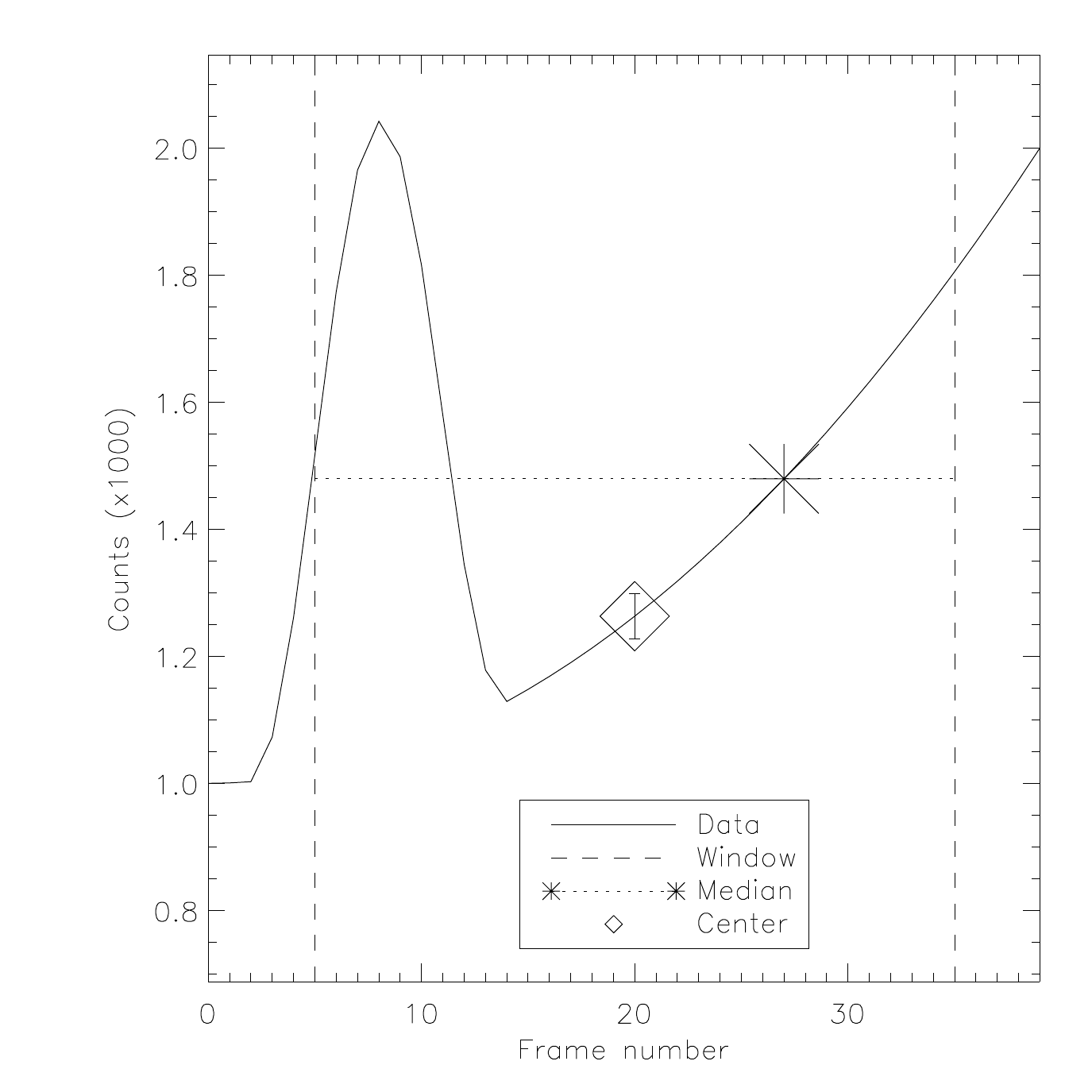}
	\end{center}
	\caption{Example illustrating response of RCM filter to a peak superimposed on a a monotonic brightening. There is an ambiguity between this signal and a dimming, and the RCM filter does not discriminate between the two.}\label{fig:peakplusmonotonic}
\end{figure}

\section{Summary \& Future Work}\label{sec:discussion}

This work finds that dim events such as the EIT wave shown can be resolved all the way to the edge of the AIA field of view. Credit for this is due in part to the excellent SNR offered by modern instruments such as AIA. They offer more sensitivity to these faint events than is often appreciated, especially when combined with temporal and spatial binning. However, good SNR alone is insufficient to resolve these faint signals when they are superimposed on a bright and (slowly) varying background. The method described here clearly separates these events from the background. Its  The use of the running median isolates such events more cleanly than the running differences most commonly used, and it does not show the gradual drift away from the reference seen in base difference images. It also allows a straightforward way to specify a maximum timescale via the width of the running window, and oscillatory behavior at short timescales can be suppressed by temporal binning. 


As mentioned in Section \ref{sec:intro}, the most similar method found in the literature is \cite{Terzoetal2011}, which compares the medians of a (per-pixel) time series to the corresponding averages. Despite being the most similar method found, it differs in some notable ways from the one presented here. Most significantly, the comparison of the median and the mean does not, in general share the same insensitivity to (non-linear) monotonic trends (a step function will register as either a negative or positive difference depending on where in the window the step lies) as the comparison of the median to the central value in the window. The method presented here also has higher time resolution, since the results will depend on whether or not the central value is affected by a peak, whereas the mean will be affected by a peak regardless of where it falls in the window. Of course, use of the mean may also give the advantage of increased sensitivity in some cases, but the above demonstrates that there are situations where center-median filtering would be preferable. 

This paper also demonstrates how to take advantage of the well-understood nature of the noise (namely, that it is primarily due to Poisson counting statistics) in AIA and similar instruments, using it to compute the significance of deviations from the running median.

The method presented here is fairly generic and straightforward, but shows EIT waves and other dynamic phenomena more clearly than the standard base and running difference techniques used in the field. This suggests that there is significant progress still to be made. Future work will consider the scientific implications of this method in more detail, and investigate a few areas which may result in significant improvement; first, the SNR is limited by the noise in the central frame of the window, but it may be possible to construct a similar method that makes full use of the entire window in searching for trends. Second, it is limited by the noise in the time series for a single pixel (or binned pixel); by incorporating spatially multi-scale decompositions of the image \citep[perhaps along the lines of][]{Stenborg_etal_ApJ2008,MorganDruckmuller_SoPh2014}, it should be possible to significantly improve the detectability of faint large-scale features (such as EIT waves) while retaining high spatial resolution. A third area is the effect of noise and bimodal distribution on the median, which tends to produce spurious brightenings or dimmings when a large change (such as a step function) is present in the window (even if the central value is not directly effected). It may be possible to compensate for this effect as well. One limitation which probably cannot be compensated in this method is the type of ambiguity illustrated by Figure \ref{fig:peakplusmonotonic}; this likely can only be resolved by a more sophisticated (and specialized) background fitting method.

The code used to produce the examples for this technique is freely available on request.
	
\begin{acknowledgements}
	The anonymous referees have made a number of helpful suggestions and comments. This paper has been aided considerably by conversation with Craig Deforest and Charles Kankelborg, among others. The author and the editor thank two anonymous referees for their assistance in evaluating this paper.
\end{acknowledgements}

\bibliography{swsc,temporalfiltering}

\end{document}